%Paper: 9201007
%From: MIRJAM@PENNDRLS.UPENN.EDU
%Date: Monday, 6 January 1992 1617-EST

\input phyzzx
\date={December 1991}    % uses PHYZZX
\Pubnum={\caps UPR--473--T, YCTP-P43-91
}

\def\to{\rightarrow}
\titlepage
\title{STATIC DOMAIN WALLS IN $N=1$ SUPERGRAVITY}

\frontpageskip=0.5\medskipamount plus 0.5 fil
\author{Mirjam Cveti\v c, Stephen Griffies}
\address{Department of Physics\break
        University of Pennsylvania\break
        Philadelphia, PA 19104--6396}

\andauthor{Soo-Jong Rey}
\address{Center for Theoretical Physics\break
Sloane Physics Laboratory, Yale University\break
New Haven CT 06511}

\abstract{
We study  supersymmetric domain walls in N=1
supergravity theories, including those
with modular-invariant superpotentials arising
in superstring compactifications.
Such domain walls are shown to
saturate the Bogomol'nyi bound of wall energy per unit area.
We find \sl static \rm and \sl reflection asymmetric \rm
domain wall solutions  of the self-duality equations
for the metric and the matter fields.
Our result establishes  a new
 class of domain walls beyond
those previously classified.
As a corollary, we define a precise notion of vacuum
degeneracy in the supergravity theories.

In addition, we found examples of global supersymmetric domain
walls that do not have an analog when gravity is turned on. This result
establishes
that in the case of extended topological defects gravity plays
a crucial, nontrivial role.}

\endpage

\chap{Introduction}

Global or local topological defects are known to arise
during symmetry breaking phase transitions if
the vacuum manifold is not simply connected. Textures,
monopoles, strings, domain walls and combinations
thereof are examples.  These objects
may have important physical implications, especially in
the cosmological context.

Inclusion of gravity in the study  of
 topological defects  is straightforward and
 usually leads to insignificant modifications to the otherwise
stable topological defects.
However, in superstring
theories, for example, gravity and other moduli and matter fields
are on an equal footing so the effects of gravity can yield
distinctly new features. With the advent
of deeper understanding of semi-classical superstring theories
in a topologically nontrivial sector, various
stringy topological defects were discovered:
stringy cosmic strings \REF\DGHR{
A. Dabholkar, G.W. Gibbons, J.A. Harvey,
F.R. Ruiz,
Nucl. Phys. \bf B340 \rm (1990) 33.} \REF\VAFA{B. Greene, A. Shapere,
C. Vafa and S.-T. Yau, Nucl. Phys. \bf B337 \rm (1990) 1.} \refmark{
\DGHR\ , \VAFA },
axionic
instantons \REF\REY{S.-J. Rey, \sl Axionic String Instantons and Their
Low-Energy Implications, \rm `Particle Theory and Superstrings', ed.
L. Clavelli and B. Harm, World Scientific Pub. Co. (1989);
Phys. Rev. \bf D43 \rm (1991) 526.} \REF\REYII{S.-J. Rey,
\sl On String Theory and Axionic Strings
and
Instantons, \rm talk given at `Particle \& Fields `91', Vancouver Canada
(1991); \sl Exact N=4 Superconformal
Field Theory of Axionic Instantons, \rm SLAC-PUB-5662 (1991).}
\REF\CHS{
 C. G. Callan, Jr. J. A. Harvey  and A. Strominger
Nucl. Phys. {\bf B359} (1991) 611.}
\REF\OT{
B. Ovrut and S. Thomas ,
\sl Instantons in Antisymmetric Tensor Theories in Four-Dimensions,\rm
 UPR-0465T, (March 1991),
 and
Phys.Lett.{\bf B}267 (1991) 227.}
\refmark{\REY\ -\OT}
as well as
 related heterotic
five-branes and other solitons
\REF\ANDY{A. Strominger,
Nucl. Phys.
\bf B343 \rm (1990) 167;
erratum \sl ibid \bf B353 \rm (1991) 565.
}
\REF\CHSII{
C. G. Callan, Jr. J. A. Harvey  and A. Strominger,
\sl Worldbrane Actions for String Solitons \rm , PUPT-1244(March 1991).}
\ \refmark {\ANDY\ , \CHS\ , \CHSII}
among others.

The above solutions were known to exist for free moduli fields, \ie
vanishing superpotential. Earlier, we have found
supersymmetric domain walls when a nontrivial
superpotential for the moduli fields does exist\REF\CQR{
M. Cveti\v c, F. Quevedo and S.-J. Rey,
Phys. Rev. Lett. \bf 63 \rm (1991) 1836.}\ \refmark{\CQR}.
Such  domain walls  are
interesting by themselves as well as
in connection to the dynamical supersymmetry
breaking mechanism in superstring theory\REF\FILQI{A. Font, L.E. Ib\'a\~nez,
 D. L\"ust and  F. Quevedo, Phys. Lett. {\bf 245B}
  (1990) 401;
S. Ferrara, N. Magnoli, T.R. Taylor and
G. Veneziano, Phys. Lett. {\bf 245B}
 (1990) 409;
H.P. Nilles and M. Olechowski, Phys. Lett. {\bf 248B}
(1990) 268;
P. Binetruy and M.K. Gaillard, Phys. Lett. {\bf 253B}
(1991) 119.}\
\REF\CFILQ{M. Cveti\v c, A. Font, L.E. Ib\'a\~nez,
D. L\"ust and F. Quevedo, Nucl. Phys. \bf B361 \rm (1991)
194.}\
\refmark{\FILQI\ , \CFILQ}.  Additionally, they
serve as a class of stringy
topological defects in which a nonzero superpotential
is essential to their existence. In this paper, we continue to elaborate on
the existence and explict solutions of supersymmetric domain walls,
completing our earlier work \refmark{\CQR}.

There are three major results of our analysis.
The first is a proof of a positive energy density theorem
for a topologically nontrivial extended object
in which the matter part has a generic nonzero superpotential.
To the best of our knowledge, the proof has never been addressed previously.
We present details of the proof for domain walls.
However, the techniques can be generalized easily to other topological
extended objects with nontrivial superpotential.

The second result is an existence proof of
\sl static \rm domain wall solutions for both the  space-time
metric and the matter field
interpolating between two supersymmetric vacua.
It is known
that the inclusion of gravity to reflection symmetric domain walls
of infinite extent and infinitesimal thickness generically admits
only \sl time-dependent \rm metric solutions\Ref\IPSK{
A. Vilenkin, Phys. Lett. \bf 133 B \rm (1983)
177; J. Ipser and P. Sikivie, Phys. Rev. \bf D30 \rm (1984) 712.}.
We show that by
allowing for a reflection asymmetric solution  interpolating
between the Minkowski and anti-deSitter space-time
the metric
and matter field can be
\sl time-independent.\rm

The last result  is that
supersymmetric domain walls
interpolate between
two vacua of different scalar potential energy: for example,
between a supersymmetric vacuum with zero cosmological constant
(Minkowski space-time)
and another with a negative cosmological constant (anti-deSitter
space-time).  This leads us to
define a notion of \sl vacuum degeneracy \rm in
supergravity theories as those vacua that are
supersymmetric.

The paper is organized as follows.
In Chapter 2 we recapitulate the results for domain walls in
global supersymmetric field theories.
We illustrate and exemplify the solutions for a double-well
potential
and for a duality invariant superpotential
which respects
$T\to1/T$ transformation motivated by superstring compactifications.
In Chapter 3  we prove
a positive energy density theorem of the topologically
nontrivial domain walls in N=1 supergravity.
The supersymmetric domain walls are found as solutions of self-dual
equations that saturate the Bogomol'nyi bound. We again exemplify
the general results using the cases discussed
in Chapter 2.
In Chapter 4  we provide physical interpretations of
the \sl static, reflection asymmetric
\rm domain walls in connection with $O(4)$ invariant bubbles of false vacuum
decay in supergravity theories.
In  Appendices A and B we
present details of the Bogomol'nyi bound calculation and the
first order self-dual equations of motion, respectively.

\chap{Global Supersymmetric Domain Walls}

In this chapter, we remind the reader of certain aspects of
domain walls arising in
global supersymmetric field theories
\REF\TOWN{ E.~Abraham and P.~Townsend, Nucl. Phys. {\bf 351}
(1991) 313.}
\refmark{\CQR\ ,\TOWN}.
Apart from its own
academic interest, the global supersymmetric case
furnishes useful insights to domain walls in
supergravity theories we will study in the next chapter.

Consider a
$N=1$ four-dimensional globally supersymmetric field theory of a chiral
superfield $\cal T$ with scalar and fermionic
components denoted by
$T$ and $\chi$, respectively.
The Lagrangian is written in terms of two non-derivative
functions: $K({\cal T }, \bar{{\cal{T}}})
$, the K\"ahler potential, and a holomorphic function
$W(\cal T)$, the superpotential. The Lagrangian is
$$
%\eqalign {L &= \int \!\! d^2 \theta d^2 \bar \theta K({\cal T },
% \bar{\cal T })
%+ \int d^2 \theta W({\cal T }) + \int d^2 \bar \theta \bar W(\bar{ \cal
% T } ) \cr
%&= \eta^{ab} K_{T \bar T}\partial_{a}T \partial_{b}\bar T
%- K^{T \bar T} |\partial_T W(T)|^2}.\eqn\lag
L = \eta^{ab} K_{T \bar T}\partial_{a}T \partial_{b}\bar T
 - K^{T \bar T} |\partial_T W(T)|^2.\eqn\lag
$$
Here, $K_{T \bar T} =1/K^{T \bar T} \equiv \partial_T \partial_{\bar T}
K(T, \bar T)$
is a positive definite metric of the space spanned by the
T fields and
$\eta^{ab} = Diag(+1,-1,-1,-1)$ specifies the flat space-time metric.
We set the fermionic component to zero in \lag\ as we are interested
in the vacuum structure.

For the theory to have a domain wall solution one needs
degenerate  isolated global minima, \ie\ the vacuum manifold
must have a nontrivial fundamental homotopy group
$\Pi_0 ({\cal{M}})\not=I$. This structure is ensured if a
discrete global symmetry is  spontaneously broken.
The absolute minimum of
the semi-positive
definite potential $V$  in eq. \lag\
satisfies
$V\equiv K^{T\bar T}
|\partial_T W|^2=0$, thus $\partial_T W  =0 $.
Such vacua preserve supersymmetry.  Therefore,
existence of \sl degenerate but isolated \rm supersymmetry preserving vacua
signals existence of a supersymmetric domain wall interpolating
between any two such vacua.

\section{Minimum Energy Solution}
We are looking
for time-independent domain wall solutions that minimize  the
energy density per unit area. Such solutions are characterized by
translational symmetry in the two directions tangential to the wall.
For such a configuration, $T = T(z)$, where $z$ is the coordinate
transverse to the wall.
The domain wall energy per unit area is
$$
\sigma \equiv
 {E \over \int dx dy} = \int^\infty_{-\infty} \!\!dz \,\,
(K_{T \bar T} |\partial_z
T|^{2} + K^{T \bar T} | \partial_T W(T)|^2).\eqn\maden
$$
This expression can be written\REF\FMVW{
P. Fendley, S. Mathur, C. Vafa and N.P. Warner, Phys.
Lett. \bf 243B \rm (1990) 257.}
\refmark{\CQR\ ,\TOWN\ ,\FMVW} as
$$
\sigma = \int^\infty_{-\infty} dz \, K_{T \bar T} |\partial_{z}T
- e^{i \theta}
K^{T \bar T} \partial_{\bar T} \bar W (\bar T)|^2
+ 2Re (e^{-i \theta} \Delta W)
\eqn\massden
$$
where $\Delta W \equiv W(T(z =\infty)) - W(T(z = - \infty))$. The
phase $\theta$ is chosen such that $e^{i \theta} = \Delta W / |\Delta
W|$, thus maximizing the cross term in \massden\ .
With such a phase choice,
the domain wall energy per unit area is bounded
by the Bogomol'nyi inequality
$\sigma
\ge |C| \equiv 2|\Delta W|$
where the complex-valued $C$ is the
topological charge or (unnormalized) \sl domain wall \rm number.

Since $\partial_T W$
is analytic in $T$, the line integral over $T$ is \sl path independent
\rm   as for
 a conservative force. The minimum is obtained only if
 the  $T$ field satisfies the first order differential
equation
$$
\partial_z T(z) = e^{i \theta} K^{T \bar T} \partial_{\bar T} \bar
W (\bar T(z))
\eqn\bogeq
$$
thus saturating the Bogomol'nyi bound\Ref\BOG{
E. B. Bogomol'nyi, Sov. J. Nucl.
Phys. \bf 24 \rm (1976) 449.}.
One can easily check that the solution
of Eq.\bogeq\
satisfies the equations of motion derived from the Lagrangian \lag\ .
In this case, $\partial_z W(T(z)) = e^{i \theta}K^{T \bar T}
|\partial_T W(T(z))|^2$, which implies the phase of $\partial_z W$
does not change with $z$. Thus, the domain wall is a mapping
from the
$z$-axis $(-\infty, \infty)$ to a \sl straight line \rm in the
superpotential plane $W \in {\bf C}^1$ connecting
two degenerate supersymmetric vacua. The domain wall is
stabilized by the topological charge $C \ne 0$.

One can understand the Bogomol'nyi bound as a consequence of
supersymmetry preserving bosonic background.
For a bosonic background,
the supersymmetry transformations of the $T$ field and its
supersymmetric partner $\chi$ (written as a four component
Majorana spinor) are
{}~\Ref\WB{J. Wess and J. Bagger,
\sl Supersymmetry and Supergravity, \rm
2nd edition, Princeton University Press 1991.}
$$
\eqalign{
\delta_{\epsilon}T& = 0, \cr
\delta_{\epsilon}\chi
 & = -\sqrt2 K^{T \bar T}(\bar W_{\bar T} P_{R}
              + W_{T} P_{L}) \epsilon
- i\sqrt2 (\partial_{a} T P_{R} + \partial_{a} \bar T P_{L})
 \gamma^{a} \epsilon
}
\eqn\sutran $$
where $P_{R,L} = {1\over2}(1 \pm i \gamma^{5})$  are
right-(left-)handed projections,
$\gamma^{5} = \gamma^{0}\gamma^{1}\gamma^{2}\gamma^{3}$,
and $\gamma^a, \, a=0,...,3$ are Dirac matrices.
For a Weyl basis of the Dirac matrices,
the constant Majorana spinor has components
$\epsilon=(\epsilon_1,\epsilon_2,\epsilon_2^*,-\epsilon_1^*)$.
A supersymmetric bosonic background is defined as a $T$ configuration
satisfying $\delta_{\epsilon}\chi = 0$.
Using \sutran\   and
keeping in mind that $T$ depends only on $z$,
one obtains
$$
\partial_z T(z) = ie^{i \alpha} K^{T \bar T} \partial_{\bar T} \bar
W (\bar T(z))
\eqn\bogeqii
$$
and a constraint on the $\epsilon$ components:
$$
\epsilon_1=e^{i\alpha}\epsilon_2^*,
\eqn\spineq
$$
in which the constant phase   $\alpha$ is undetermined.
If one
requires  the configuration to
minimize the energy functional, then
$ie^{i \alpha} = e^{i \theta} = \Delta W / |\Delta W|$.  A general
$N=1$ supersymmetry transformation is specified by a four-component
Majorana spinor $\epsilon$ with
four real degrees of freedom. However, \spineq\
reduces  these four  degrees  of freedom to two.
Thus the domain wall solutions realize half of the
\sl $N= 1$ \rm
supersymmetry linearly, while the other half
nonlinearly as two fermionic zero modes.

The topological charge $C$  can be understood in terms of the
central extension of extended supersymmetry algebra.
Since the domain wall has a planar symmetry,
we can `compactify' the system to
$1+1$ dimensions. The four-dimensional supersymmetry charge
algebra becomes an $N=2$ supersymmetry
algebra admitting a
central charge.\foot{In four dimensions,
$N=1$ supersymmety does not allow for a central extension.
After a `dimensional reduction', the algebra of
supercharge \sl density \rm admits an $N=2$ supersymmetry which
allows for a central charge.}

The supersymmetry charges (in four component notation)
$Q_{\alpha}$
generate field
transformations:
$$
\delta_{\epsilon}\chi_{\alpha}  = -i\{\chi _{\alpha}, \bar Q_{\beta}\}
\epsilon_{\beta}.\eqn\sutrans
$$
Using eq. \sutran\ ,
we find the supersymmetry charge \sl density \rm
 $Q_{\alpha}$
by inspection:
$$
Q = i\sqrt2 \int dz (W_{T}P_{L} +
    \bar W_{\bar T}P_{R})\gamma^{0} \chi
+ \sqrt2 \int dz      K_{T \bar T}(\partial_{3}T P_{L} +
\partial_{3} \bar T P_{R} ) \gamma^{3}\gamma^{0} \chi.
\eqn\qexpl
$$
The algebra satisfied by $Q$ is the Poincar\'e $N=2$
supersymmetry algebra with a central extension due to
the nontrivial topology of the vacuum manifold\Ref\WO{
E. Witten, D. Olive,  Phys. Lett. \bf 78B \rm (1978) 97.}.
In the wall's
rest frame, the supercharge density algebra is
$$
\bar{\epsilon}_{\alpha}\{Q_{\alpha},\bar Q_{\beta}\}\epsilon_{\beta}
=
2\epsilon^{\dagger}_{\alpha}(\sigma\delta_{\alpha \beta}
+ 2i\gamma^{3}_{\alpha \beta}\Delta(ReW
+ \gamma^{5} ImW))\epsilon_{\beta}
\equiv
2\epsilon^{\dagger}(\sigma
+ C)\epsilon
\eqn\comrel
$$
In deriving \comrel\ we used  the
explicit form \qexpl\ of $Q_\alpha$ and the
canonical
equal-time anti-commutation relation for spinors in a non-trivial
K\"ahler background:
$\{\chi_{\alpha}(z), \bar \chi_{\beta}(z')\}  =
(\gamma^{0})_{\alpha \beta} K^{T \bar T} \delta( z-z')$.

Clearly, the first term on the right hand side of eq. \comrel\
corresponds to the  domain wall
energy density (defined in eq. \maden\ ))
and the second term,\foot{ $C$ in this equation is a
matrix.  We use the same symbol to represent its eigenvalue
in the bound $\sigma \ge |C|$}
which is topological,  corresponds to
the central charge of $N=2$ supersymmetry algebra in $1+1$ dimensions.

In view of the above analysis of the supersymmetry transformations
and the corresponding supersymmetry charge-density algebra
with central charge, we can rewrite the
Bogomol'nyi bound  in terms of a semi-positive definite expression
which is
zero if and only if the supersymmetry transformation \sutran\
vanishes. Such an expression will
be useful in the local case as well.

Consider the bilinear
$\Gamma^{a} = K_{T \bar T} \overline{\delta_{\epsilon}\chi} \gamma^{a}
\delta_{\epsilon}\chi$.\foot{
We previously assumed
the Majorana spinor  $\chi$ to be anticommuting.  This
would in turn imply that
$\Gamma^{a}$ vanishes identically.
However, the anticommuting spinor parameter $\epsilon$
can be expressed as a linear combination
$\Sigma\rho^{m}\epsilon^{m}$,
($m=1,\ldots 4$). Here,
 $\rho^{m}$ is an odd element of the
Grassman algebra ($\rho^{m}\rho^{n} = -\rho^{n}\rho^{m}$), and the
$\epsilon^{m}$ are a set of independent \sl commuting \rm
Majorana spinors.  These considerations can be generalized to
local supersymmetry parameters $\epsilon(x)$\Ref\hull{
C.M. Hull, Commun. Math Phys. \bf 90 \rm (1983)
545.} as well.
Therefore, $\Gamma^{a}$ is interpreted as
a bilinear of \sl commuting \rm Majorana spinors
$\epsilon^{m}$.  With this  proviso, $\Gamma^0$ is semi-positive
definite.}
In line with the supercharge \sl density \rm introduced in
eq.\qexpl\ ,
we interpret
$\Gamma^a$ as a density.
Integration of $\Gamma^a$ over
a space-like
hypersurface (here the z-axis\foot{$\Gamma^{a}d\Sigma_{a}=
\Gamma^{0}dz$})
yields
$$
\eqalign{
\int dz \Gamma^{0}
&=
2\epsilon^{\dagger} \int dz ( K^{T \bar T}|W_{T}|^{2}
  + K_{T \bar T} \partial_{z}T \partial_{z}\bar T
  + 2i \gamma^{3} \partial_{z}(ReW + \gamma^{5}ImW) ) \epsilon\cr
&=
2\epsilon^{\dagger}( \sigma + 2i \gamma^{3} \Delta(ReW + \gamma^{5}
ImW))\epsilon \equiv
\bar{\epsilon}_{\alpha}\{Q_{\alpha},\bar Q_{\beta}\}\epsilon_{\beta}
\ge 0.
}
\eqn\formin$$
Since $\int dz\Gamma^{0} \equiv \int dz
K_{T \bar T}
\delta_{\epsilon}\chi^{\dagger}
\delta_{\epsilon}\chi \ge 0$, eq. \formin\
vanishes if and only if
$\delta_{\epsilon}\chi =0$.
{}From the second line of eq. \formin\ we find that the domain
wall energy density
$\sigma$ satisfies the
Bogomol'nyi bound
$\sigma \ge 2|\Delta W|\equiv |C|$ which is saturated if and only if
$\delta_{\epsilon}\chi = 0$.

\section{Examples}

We illustrate the above general discussions with some examples.

\noindent {\it 1. Double-Well Potential}\hfill\break
The first example is a
field theory which allows for a spontaneous breakdown
of a discrete ${\bf Z}_2$ symmetry.
We choose a minimal
K\" ahler potential:
$$K=T\overline{T}
\eqn\kaft
$$
and a cubic superpotential:
$$
W=({1 \over 3} T^3-a^2 T)
\eqn\suft
$$
where $a>0$.

The scalar potential
$$V=K^{T\bar T}|\partial_T W|^2= |T^2-a^2 |^2
\eqn\poft
$$
has two global disconnected minima $T_{\pm}=\pm{a}$.
Specifying the solution which interpolates from $T_{-}$ at
$z=-\infty$ to $T_{+}$ at $z=+\infty$ implies $e^{i\theta}=-1$
and the
self-dual equation \bogeq\  is
$\partial_z T(z)=-(\bar T^2-a^2 )$ which interpolates along
 the real value of $T(z)$ between the two minima $T_{\pm}$.
   The solution for  $T$ is the
familiar kink: $T(z)=a$tanh$(za)$.  The scalar potential
for $T=\bar{T}$ and the kink
the solution $T(z)$  for $a^{2}= 0.5$ are
displayed  in figs. 1 and
2, respectively.
We find the topological charge $2[ W(T(z=+\infty))-W(T(z=-\infty))] =
(-8/3)a^{3}$.

\noindent {\it2. Modular Invariant Potential}\hfill\break
Now consider a
field theory invariant under a modular transformation on the
scalar field $T$.  This example is
motivated by low-energy
effective Lagrangians of certain superstring compactifications.

A generalized field-space duality is characterized by a non-compact
discrete group
$PSL(2,{\bf Z})=  SL(2,{\bf Z})/ \bf{Z_2}$
specified by the linear fractional transformations:
$$ T\rightarrow{aT-ib\over icT+d}\,\, \ a,b,c,d\in{\bf Z}\ \ ,
  \ \ ad-bc=1.
\eqn\frt
  $$
This is an exact symmetry of (2,2) string vacua
based on orbifold compactifications  not only at the string tree
level\Ref\LMN{
J. Lauer, J. Mas and H.P. Nilles, Phys. Lett. {\bf 226B}
 (1989) 251;
W. Lerche, D. L\"ust and N.P. Warner,  Phys. Lett. {\bf
231B}
 (1989) 417;
E.J. Chun, J. Mas, J. Lauer and H.P. Nilles, Phys. Lett. {\bf
233B}  (1989) 141.}
but also at the world-sheet nonperturbative level.  This result
was further supported by genus-one
threshold calculations \Ref\DKLI{L. Dixon, V. Kaplunovsky and J. Louis,
Nucl. Phys. \bf 355B \rm (1991) 649.}
 which in turn  specify
the form of the gaugino condensate\refmark{\FILQI}.
The complex
modulus field $T$ corresponds to the
compactification  dilaton (internal radius of compactified space)
and the internal  axion field.
The $T$ field has no potential at the string tree level
as well as to all orders in string perturbation.
On the other hand
it is known that nonperturbative stringy effects such as
gaugino condensation \Ref\DIN{
J.P. Derendinger, L.E. Iba\~nez and H.P. Nilles,
Nucl. Phys. {\bf  267B}
(1986) 365.}
 and  axionic string instantons\refmark\REY\ give rise to
nonperturbative superpotentials. Such potentials in turn preserve
duality symmetry\REF\FLS{
S. Ferrara,
 D. L\"ust, A. Shapere and S. Theisen,
Phys. Lett. \bf 225B \rm
 (1989) 363.}\
\refmark{\FLS\ ,\FILQI\ ,\CFILQ}.
Therefore, we  study stringy domain walls of N=1 supersymmetric
four-dimensional  superstring vacua by taking
into account the modular invariant superpotential
of the $T$-modulus field.

Consider a global supersymmetric theory with $
PSL(2,{\bf Z})$ invariance and scale-invariant K\" ahler potential:
$$K=-3ln(T+\overline{T}).\eqn\kadual$$
The superpotential
$W$ is a modular invariant (weight zero) function
of $PSL(2, \bf Z)$ defined over the fundamental domain $\cal D$ of
the T-field(see fig. 3).
The most general form of the superpotential is a rational
polynomial $P(j(T))$
of the modular-invariant function $j(T)$.\Ref\SCH{
B Schoeneberg,
\sl Elliptic Modular Functions, \rm Springer-Verlag 1974.}
The simplest choice for a
modular invariant superpotential is:
$$W(T) = j(T) .\eqn\sdual$$
We have set the string tension $\alpha^\prime$ to one.
\foot{In general, the superpotential has an overall scale
depending exponentially on the dilaton field. The scale is
set to one here.}
In the fundamental domain (see fig. 3), the scalar potential
has two isolated
degenerate minima at $T=1$ and $T = \rho \equiv e^{i \pi/6}$. At
these fixed points, $j(T=1) = 1728$ and $j(T = \rho ) = 0$.
In the $e^{K/2}W$ plane the geodesic is a straight  real  line
interpolating between $W=0$ and $W=1728$.
In the complex $T$-plane the geodesic is $T=e^{i\varphi(z)}$;
\ie\ it lies along the boundary of the fundamental domain (see fig. 3).
This fact follows from the  property of
the $j(T)$ function
{}~\refmark{\SCH} that
values of $T$ yielding
a real $j(T)$ lie along the boundary of the fundamental
domain in the $T$-plane.

The scalar potential along the geodesic is depicted in fig. 4,
while the numerical solution  for $T(z)$ is given in fig. 5.
\foot{To obtain the solution for $T(z)$ we scaled the
superpotential by a factor $\Omega = 2 \times 10^{-5}$.  We found
this scaling necessary as the modular covariant functions
change by many orders of magnitude during the numerical initegration
routine.  We consider this scaling merely a lack of computational power.
However, recall the overall scale of the superpotential
depends exponentially on the dilaton field.}
The domain wall mass per unit area is $\sigma = 2 \times
1728 $.

We note that a naive
application of the topological charge
implies that the domain wall solutions between the
minima connected by $PSL(2, {\bf Z})$ mappings have zero energy
since
$W$ has the same value at these points. However, one can show that
in the fundamental domain $\cal D$ there are always \sl at least \rm
two degenerate minima with different values of the superpotential $W$,
hence the domain wall mass per unit area is nonzero. The energy density
of the domain walls interpolating between the minima connected by
$PSL(2, {\bf Z})$ transformations are determined by
taking the path through all the minima in between.
`In between' in the complex $T$-plane is unambiguous because we
know the path in the $W = j$ plane is the straight line along the
real axis.  Therefore, the path in the $T$ plane is along
the boundary of its fundamental domain.
In this case, the phase $\theta$ is adjusted
between adjacent minima to maximize the topological
term in the Bogomol'nyi self-duality equation. Therefore,
the mass per unit area of the domain wall
interpolating between $T=e^{i \pi/6}$ and $T=e^{-i\pi/6}$ is
$\sigma = 2 \times 2 \times 1728 $.

\chap{Supersymmetric Domain Walls in N=1 Supergravity}

In this section we investigate domain walls arising in
supergravity theories.  Specifically, we study $N=1$
supergravity theories in four dimensions with
a nontrivial superpotential.  As we shall see, the
topological charge of the global case can be generalized to
incorporate
the effects of gravity.  In addition, the
presence of gravity is seen to allow for domain walls interpolating
between supersymmetric vacua which are not degenerate in the
usual sense.

Consider an $N=1$ locally supersymmetric theory with one chiral
matter superfield $\cal T$.
We can
straightforwardly generalize our results to multi-matter superfield
cases.
The bosonic part of the $N=1$ supergravity Lagrangian
is~\refmark{\WB}\foot{We do not choose
the commonly used  K\"ahler gauge
which introduces the potential function\refmark{\WB}
$G(T,\bar T) = K(T,\bar T) + ln|W(T)|^{2}$, since it is not
adequate for situations in which the superpotential is allowed to vanish.}
$$
e^{-1}L = -{1 \over 2}R + K_{T \bar T}g^{\mu \nu}\partial_{\mu}\bar T
\partial_{\nu}T - e^{K}(K^{T \bar T}|D_{T}W|^{2} - 3|W|^{2})
\eqn\localL$$
where
$e = |detg_{\mu \nu}|^{1 \over 2},
K(T, \bar T) =$ K\"ahler potential and $D_{T}W \equiv e^{-K} (\partial_T
e^K W)$.
\foot{We use
the conventions: $\gamma^{\mu}=e^{\mu}_{a}\gamma^{a}$ where
$\gamma^{a}$ are the flat spacetime Dirac matrices satisfying
$\{\gamma^{a},\gamma^{b}\}=2\eta^{ab}$; $e^{a}_{\mu}e^{\mu}_{b}
= \delta^{a}_{b}$; $a=0,...3$; $\mu=t,x,y,z$; $\overline{\psi} =
\psi^{\dagger}\gamma^{t}$; $(+,-,-,-)$ space-time signature;
and dimensions such that
$8\pi G \equiv 1$.}

In order to have stable domain wall solutions,
topological arguments imply that the degenerate vacua
be disconnected;
\ie\  the  fundamental homotopy group
of the vacuum manifold satisfies
$\Pi_0({\cal{M}}) \ne I$. Thus
one must have isolated
vacua of the matter potential. However, inclusion of gravity
will turn out to play an important role in removing the
constraint  that the isolated minima of the potential have to
be degenerate.  We shall see
the notion
of degenerate vacua will be defined as
\sl supersymmetry preserving vacua. \rm

Supersymmetry preserving minimum of the  potential
in \localL\ satisfy
$D_{T}W=0$. This in turn implies (see eq.\localL\ )
 that the supersymmetry preserving vacua have
 either  zero cosmological constant (Minkowski space-time)
 when $ W=0$, or negative  cosmological constant
 $-3e^K|W|^2$  (anti-deSitter space-time) when
$W \not=0$.

\section{ADM Mass Density}
In the following
we will obtain a lower bound on the
mass density of domain walls
%which interpolate between the
%the isolated supersymmetric vacua.
in this theory.
The bound
is a generalization of the global result.
We employ the results of
Refs.~\Ref\WITTEN{
E. Witten, Comm. Math. Physics,
\bf 80 \rm (1981) 381.}
and
{}~\Ref\NESTER{
J.M. Nester, Phys. Lett. \bf 83A \rm (1981) 241.}
which addressed the positivity of the ADM
mass in general relativity, as well as certain
generalizations to  anti-de Sitter backgrounds
{}~\Ref\GHW{
G.W. Gibbons, C.M. Hull, N.P. Warner,
Nucl. Phys. \bf B218 \rm (1983) 173 ;
C.M Hull, Nucl. Phys \bf B239 \rm (1984) 541.}.
The ADM mass\Ref\RELA{C. Misner, K. Thorne and
J. Wheeler, \sl Gravitation, \rm 1973; R.M. Wald, \sl General Relativity,
\rm 1984.}
for spatially infinite objects is
not well-defined\Ref\DES{S. Deser, Class. Quantum. Grav. \bf 2 \rm
(1985) 489.}.  However, as a weaker requirement, we will assume that
the ADM procedure is valid for the mass per unit area rather than the
mass of the domain wall.

Consider the supersymmetry charge density
$$
Q[\epsilon'] = \int_{\partial \Sigma} \bar{\epsilon'}
\gamma^{\mu \nu \rho} \psi_{\rho} d\Sigma_{\mu \nu}
\eqn\localcharge$$
where $\epsilon'$ is a commuting Majorana spinor,
$\psi_{\rho}$ the spin $3/2$ gravitino field, and
$\Sigma$ a spacelike hypersurface.
We take a supersymmetry variation of $Q[\epsilon']$ with respect to
another commuting Majorana spinor $\epsilon'$
$$
\eqalign
{
\delta_{\epsilon} Q[\epsilon']& \equiv \{Q[\epsilon'],
\bar{Q}[\epsilon]\}  \cr
%&= \int_{\partial \Sigma} d \Sigma_{\mu \nu} \bar \epsilon'\gamma^{\mu
%\nu \rho   }
%\hat\nabla_{\rho} \epsilon \cr
&= \int_{\partial \Sigma}N^{\mu \nu} d\Sigma_{\mu \nu}
= 2\int_{\Sigma}\nabla_{\nu}N^{\mu \nu} d\Sigma_{\mu}  \cr
}
\eqn\localchargevariation$$
where
$N^{\mu \nu} = \bar \epsilon'\gamma^{\mu \nu \rho}
\hat\nabla_{\rho} \epsilon $ is a generalized
Nester's form\refmark\NESTER\ . Here
$\hat\nabla_{\rho}\epsilon \equiv
\delta_{\epsilon}\psi_{\rho} =
[2\nabla_{\rho} + ie^{K \over 2}(WP_{R} + \bar{W}P_{L})\gamma_{\rho}
 - Im(K_{T}\partial_{\rho}T)\gamma^{5}]\epsilon$ and
 $\nabla_{\mu}\epsilon = (\partial_{\mu}
  + {1\over2}\omega^{ab}_{\mu}\sigma_{ab})\epsilon$.
 In \localchargevariation\
the last equality follows
from Stoke's law.

We consider an Ansatz for
the space-time metric
$ds^{2} = A(z,t)(dt^{2} - dz^{2}) +\break
 B(z,t)(-dx^{2} - dy^{2})$
characteristic of space-times with a domain wall where $z$ is
the coordinate transverse to the wall.
However, we do not assume \sl a priori \rm
that the metric is symmetric about the plane $z=0$. Nor do we
assume a particular behavior of $A$ and $B$ at
$|z|\rightarrow\infty$ except that the asymptotic metric satisfies
the vacuum Einstein equations  with a zero or negative cosmological
constant.

We are concerned with supercharge \sl density \rm and thus
insist upon only $SO(1,1)$ covariance in the $z$ and $t$ directions.
This in turn implies
that the space-like
 hypersurface
$\Sigma$ in eq.\localchargevariation\
is the $z-$axis
with measure
$d\Sigma_{\mu} = (d\Sigma_{t},0,0,0) =
|g_{tt}g_{zz}|^{1\over 2}dz$.  The boundary
$\partial \Sigma$ are then the
two asymptotic points $z\rightarrow \pm\infty$.
Technical details in obtaining the explicit form of
eq.\localchargevariation\
are given in the Appendices A and B. Here we only quote the final
results.

  The volume integral
yields:
$$
  2\int_{\Sigma}\nabla_{\nu}N^{\mu \nu} d\Sigma_{\mu}=
%\delta_{\epsilon} Q[\epsilon'] =
\int_{-\infty}^{\infty}
[-\delta_{\epsilon'}\psi^\dagger_i g^{ij} \delta_{\epsilon}\psi_j +
K_{T \bar T}\delta_{\epsilon'}\chi^\dagger \delta_{\epsilon}\chi]dz
\eqn\volumeintegral$$
where $\delta_{\epsilon}\psi_{i}$ and $\delta_{\epsilon}\chi$
are the supersymmetry variations of the fermionic fields in the
bosonic backgrounds.
Upon setting $\epsilon' = \epsilon$ the expression \volumeintegral\
is a positive definite quantity which in turn (through
 eq.\localchargevariation\ )
yields  the bound
$\delta_{\epsilon} Q[\epsilon] \ge 0.$

Analysis of the surface
integral in \localchargevariation\
yields two terms: $(1)$ The ADM mass density of configuration,
denoted $\sigma$ and
 $(2)$ The topological charge density,
denoted $C$ (see Appendix A).

Positivity of the volume integral
translates into the bound
$$
\sigma \ge
 |C|
\eqn\localbound$$
which is saturated iff $\delta_\epsilon Q[\epsilon]=0$. In this
case   the  bosonic backgrounds are supersymmetric, \ie\
they satisfy
$\delta\psi_{\mu} = 0$   and
$\delta\chi = 0$ (see eq.\volumeintegral\ ).\foot{
$\delta_\epsilon Q[\epsilon] = 0$ seems to only require
$\delta_{\epsilon}\psi_{i} = 0$ with
 $i \ne t$.  However, in order for
$\delta_{\epsilon}\psi_{i} = 0$
for an \sl arbitrary \rm space-like hypersurface,
one in fact requires
$\delta_{\epsilon}\psi_{\mu} = 0$
for $\mu = t,x,y,z$ ~\refmark\WITTEN .}
Such configurations saturate the previous bound, as in the global
case (cf. \formin\ ).

\section{Self-Dual Equations}
We now concentrate on solving for the space-time metric
and matter field configuration in the supersymmetric
case.  This calculation involves an analysis of
the first order equations
$\delta_{\epsilon} \psi_\mu = 0$ and $\delta_{\epsilon}
\chi = 0$ which are given in Appendix B.
\foot{We call these equations \sl self-dual \rm in the sense
as the first order differential equations whose square gives
the classical equations of motion.}
The self-dual equation
for the matter field $T(z)$ follows from
$\delta_{\epsilon} \chi = 0$  as in the global case:
$$
\partial_z T(z) = ie^{i \theta}\sqrt{A}
e^{K\over 2}K^{T \bar T}D_{\overline{T}}\overline{W}
\eqn\localbogeq
$$
with a constraint on the $\epsilon$-spinor:
$$
  \epsilon_1=e^{i\theta}\epsilon_2^*.
\eqn\localspineq
$$
Unlike
the global supersymmetric case, however, the yet-undetermined phase
$e^{i\theta}$ is in general \sl not \rm a constant but a space-time
coordinate-dependent function.

Since we wish to define the ADM mass per unit area
of the domain
wall unambiguously, we
look for a \sl time-independent \rm metric solution.
In  Ref.~\refmark\IPSK\
domain walls  in general relativity were studied.
It was concluded that,
even though
the energy-momentum tensor of the domain wall is time-independent, the
assumption of a reflection symmetric asymptotic
Minkowski space on both sides of the
domain wall
render the space-time metric
time-dependent generically
(unless one takes a special value of mass to tension ratio that is not
realized by generic field theory examples).
With no assumed reflection symmetry of the space-time metric,
\sl a priori \rm
one cannot say if there exist nontrivial time-independent
domain wall solutions.
We note that in order for our assumption of
time
independence of the $T$-field to be consistent with the Bogomol'nyi
equation \localbogeq\ ,
the metric component $A$ \sl must \rm be time-independent as well.

Self-dual equations for the metric components, following from
$\delta_{\epsilon}\psi_t
= \delta_{\epsilon} \psi_x = 0$ (see appendix B)
%along with \localbogeq\ and \localspineq\ ,
are
$$
\partial_{z}lnA =
\partial_{z}lnB =
2(ie^{-i\theta})\sqrt{A}e^{K\over 2}W.
\eqn\metriceq
$$
Since the metric functions
$A$ and $B$ are real, the phase $e^{i\theta}$ is required to meet
a local constraint
$$
W = -i\zeta e^{i\theta}|W|
\eqn\thetaconstraint$$ where $\zeta = \pm$.
Assuming continuity, $\zeta = \pm$ can
change
only at points where
 $W$ vanishes.

Additionally,
$\delta_{\epsilon} \psi_z=0$ yields the differential equation
for the $z$ dependent phase $\theta$:
$$
\partial_{z} \theta = - Im(K_{T}\partial_{z}T).
\eqn\otherthetaconstraint$$
Consistency of \localbogeq\ , \metriceq\ and \otherthetaconstraint\
with
\thetaconstraint\ leads to following sufficient conditions for the
existence of a static supersymmetric domain wall:
$$
\eqalign{
\partial_z T(z) &= -\zeta \sqrt{A}|W|
e^{K\over 2}K^{T \bar T}
{D_{\overline{T}}\overline{W}\over \overline{W}},\cr
\partial_{z}lnA &=
\partial_{z}lnB =
2\zeta\sqrt{A}|W|e^{K\over 2}, \cr
Im&(\partial_{z}T{D_{T}W\over W}) = 0.}
%\epsilon_1 &= e^{i\theta}\epsilon_2^*.}
\eqn\summary$$

We now comment on these three equations.

(i) The first equation in \summary\ is a local
generalization of the
global result \bogeqii\ . It is evident that $\partial_z T(z)
\rightarrow 0$ as one approaches the supersymmetric minima,
\ie\ $D_T W=0$, thus indicating
a domain wall configuration.

(ii)
The second equation  in \summary\ , \ie\ the
 equation for the  metric, implies that we can always
rescale the space-time coordinates to bring $A = B$. Thus, our
metric ansatz is reduced to a class of conformally flat metrics
with $z$-dependent conformal factor.
The asymptotic behaviour of the metric depends on whether
the supersymmetric vacuum is Minkowski ( $|W_{\pm \infty}|=0$) or
anti-deSitter ($|W_{\pm \infty}|\ne 0$).
In the first case the metric equation gives
$A\rightarrow const$,
while in the second case $A\rightarrow const'/z^2$,
which are the proper asymptotic behaviours in Minkowski
and anti-de Sitter space-times, respectively.\foot{
Note that the asymptotic metric
$\propto z^{-2}$ yields a negative cosmological constant,
thus correponding to an asymptotic anti-deSitter space.
Namely, we can write the cosmological constant
as $\Lambda \equiv - 3({ \partial_{z}lnA \over \sqrt{A} })^{2}$.
For $\sqrt{A}=1/(-\zeta |W_\infty|e^{K_\infty \over 2} z)$,   $\Lambda$
is found to be a constant, negative value
$-3   |W_\infty |^2e^{K_\infty}$.
Note also that
the proper distance
$d(z) = \int^{z} A^{1\over2}(z')dz'$ grows as ln$z$ since $A \propto
z^{-2}$.
This coordinate  system therefore
 completely covers the space-time.}

Another comment regarding the choice of $\zeta\equiv\pm 1$
is in order.
In the case when $W_{\pm\infty}\not=0$, \ie\
anti-deSitter
space-time on both  sides of the domain wall,
the metric approaches the value
$\sqrt A=
1/(-\zeta |W_{\pm\infty}|e^{K_{\pm\infty}\over 2}z)$
with $z\to\pm\infty$.
Note that for a real value of the metric $A$ one
has to have $\zeta=\pm 1$ for $z\to \mp\infty$.
A consequence of this observation is that when one interpolates
between the two anti-deSitter supersymmetric vacua
the value of the $\zeta$ parameter has to be different  at
each minimum in order to avoid a   singular behaviour
of the metric. Additionally,
the change of the  $\zeta$ parameter
from $+1$ to $-1$ can take place iff $W$
passes through zero along the path from $z=-\infty$ to $z=+\infty$
(see eq.\thetaconstraint ).
If this is not the case, the metric of the domain wall
interpolating between the two anti-deSitter vacua incures a
singularity  on either side of the domain wall.

(iii)
The third equation in  \summary\
 describes a \sl geodesic path \rm
between two supersymmetric vacua
in the supergravity potential space $e^{K/2} W \in \bf C$ when
 mapped from the
$z$-axis $(-\infty, + \infty)$.
Here, we
 would like to contrast the geodesic equation in \summary\
with the  geodesics in the
global supersymmetric case. In the global case the
geodesics
are straight lines in the $W-$plane (see discussion after eq.
\bogeq\ ).
On the other hand, the local
 geodesic equation in
 the limit $G_{N} \to 0$ (global
limit of the local supersymmetric theory)
leads to the geodesic equation
$Im({\partial_{z}W\over W})\equiv
\partial_z \vartheta = 0$
where $W$ has been written
as $W(z) =|W|e^{i\vartheta}$.
This in turn implies that
 as $G_N\rightarrow 0$ the geodesic equation
reduces to the constraint that  $W(z)$
has to be \sl a  straight line passing through the
origin\rm ;
\ie\ the phase of $W$ has to be
 \sl constant mod $\pi$ \rm.

The above observation in turn implies that
the introduction of gravity
imposes a strong constraint on the type of
domain wall solutions. In particular,
domain wall solutions in the global case interpolating
between vacua
in the $e^{K/2}W$ plane
that do not
lie along a straight line passing through the origin
\sl do not \rm have  an analogous solution in the local case. \foot{
We discuss a case of $Z_3$ domain walls as an illustrative example of
this phenomenon later.}

We would now like to turn to the discussion of the
energy density of the above minimal energy solution.
In Appendix B we  find an explicit expression for the ADM mass density
of the supersymmetric domain wall configuration
$$
\sigma = |C| \equiv 2 |(\zeta|We^{K \over 2}|)_{z=+\infty}
-(\zeta|We^{K \over 2}|)_{z=-\infty}| = 2|\Delta (\zeta|W e^{K\over
2}|)|
\eqn\saturation$$

Again note that for the domain wall
solutions interpolating between
two anti-deSitter minima, with $W$ passing through
zero,
the value of
$\zeta$ takes $+1$ at $z=-\infty$ and $-1$ at $z=+\infty$.

\saturation\ constitutes a
generalization of the global case \formin\ .
Furthermore, the supersymmetry charge density algebra introduced in
\localchargevariation\ is identical to the global algebra
\comrel\ after the substitution $W \rightarrow e^{K\over 2}W$.
Note, however, the topological charge in the local case
$C=2\Delta(\zeta|We^{K\over2}|)$ is real whereas in the
global case $C = 2\Delta W$, which is complex (see  Appendix A).
Actually, this result
is a natural generalization since the K\"ahler potential and the
superpotential in the global supersymmetric field theory should
combine into the supergravity potential in K\"ahler gauge
as a real section
$G = 2 Re ({1 \over 2} K + \ln W)$.
Additionally,
the reality
of the topological charge in the local theory is consistent
with the geodesic constraint.  Correspondingly, the local
Bogomol'nyi bound \localbound\ has the
$G_{N} \to 0$ limit $\sigma \ge 2|\Delta(\zeta|W|)|$ which is
equal
$\sigma \ge 2|\Delta(\zeta W)|$ only if the phase of $W$ is the
same in both vacua.

Another important comment is in order. It follows from
\saturation\ that there exists
{\it no static domain wall solution saturating the Bogomol'nyi bound
that interpolates between
two supersymmetric vacua with
zero cosomological constant.} In this case
$W(+\infty)=W(-\infty)=0$
and thus there is no
energy associated
with such a domain wall since $|C|\equiv 0$.
This result is in agreement with the results
of Ref.\refmark{\IPSK}.
Namely, these authors have not found any static solution
for infinitely thin
reflection symmetric domain walls in the
presence of gravity with asymptotic
Minkowski space-times on both sides of the domain wall.
In our study
 of the general
supersymmetric domain walls interpolating
between two
isolated supersymmetric vacua
with
Minkowski space-times
we have found that static solutions do not exist either; \ie\
there is
no topological charge associated with the Bogomol'nyi type solution.

\section{Examples}
In the following, we explore some examples
and solve for $A = B$ and $T$
explicitly.

\noindent {\it 1. Double-Well Potential}

The first example is the the minimal ${\bf Z}_{2}$ field theory:
$$ \eqalign{
K =&T\overline{T}\cr
 W=&{1 \over 3} T^3 - a^2 T.}\eqn\toysuper$$
 The matter scalar potential
%\foot{
%In order to keep track of gravitational corrections in the potential
%we explicitly introduced back
%$G_N\equiv 1/8\pi M_{pl}^2$.}
 $V \equiv e^{    K}(K^{T\bar T}
|D_{T}W|^{2} - 3    |W|^{2})$
is minimized when the K\"ahler covariant derivative
on $W$ vanishes:
$D_{T}W \equiv \partial _T W+K_T W
= 0$.  Here $K^{T\bar T}=1/\partial_T\partial _{\bar T}  K$ and $K_T=
\partial _T K $.
These vacua are supersymmetric.  We find that the line
$T(z)=\overline{T(z)}$ for the domain wall configuration satisfies the
geodesic equation
$Im(\partial_{z}T{D_{T}W\over W}) = 0$.
 The scalar potential  along the geodesic
 is given in fig. 6
for $a^{2}=0.5$.
The  geodesic has  the property that
$ie^{-i\theta (z)} = 1$.  The coupled self-dual equations
for $T$ and $A$ are
$$
\eqalign{
\partial_{z}T &=
-\sqrt{A}e^{T^{2}\over2}({1\over 3}T^{4}+(1-a^{2})T^{2}-a^{2}) \cr
\partial_{z}A &= 2A^{3\over2}e^{T^{2}\over2}T({1\over 3}T^{2}-a^{2})
}
\eqn\localtoy$$

As $\partial_{z}T$ is symmetric under $z \rightarrow -z$ and
$T \rightarrow -T$, we expect $T(z) = -T(-z)$. This is the
form of a kink centered at $z=0$.
Similarly, since $\partial_{z}A$ is anti-symmetric under
$z \rightarrow -z$, then $A(z) = A(-z)$.
Numerical solutions of \localtoy\  for
$a^{2}=0.5$ is given in figures 7 and 8 for
the field $T(z)$ and the metric $A(z)$, respectively.
The metric $A(z)$ falls
off as $z^{-2}$ as $|z| \rightarrow \infty$.  This behavior can be
extracted analytically
from an asymptotic analysis of the expression
for $\partial_{z}A$ with $K$ and $W$ approaching
nonzero constants at $\pm \infty$.

Note the two supersymmetric vacua are degenerate with
negative
cosmological constant  $ -3|W(T_{\pm})|^{2}e^{T^{2}_{\pm}\over2}$
and are thus anti-deSitter space-times (see fig. 6).
Additionally, the energy associated with the topological charge is
non-zero
and equals
$2(W(T_+)e^{T^2_+\over 2}-W(T_-)e^{T^2_-\over 2})
= 4W(T_+)e^{T^2_+\over 2}$.
Since $W$ passes through zero, the value of $\zeta$
is $+1$ on the $z<0$ side of the domain wall and $-1$ on the
$z>0$ side.

\noindent {\it 2. Modular Invariant Potential}

In the modular invariant theory, the K\" ahler potential
and the simplest form of the
superpotential are:
$$
\eqalign{
K &= -3ln[(T + \overline{T})|\eta(T)|^{4}]  \cr
W &= j(T)  }
\eqn\localstringy$$
Here, $\eta(T)$ is the Dedekind eta function, a modular form of weight
$1/2$
and $
\hat G_2 = \break -{4\pi}\partial_T \ln \eta - 2\pi/(T+\bar T)$
is the non-holomorphic Eisenstein function of weight
 two\refmark{\SCH}.
 Note
$\eta$ is regular everywhere
in the fundamental domain, while $\hat G_2$ has two zeros, one at
$T=e^{i\pi\over 6}\equiv\rho$ and another at $T=1$.
Note also  that in the local case in order to ensure modular
invariance one had to modify the
K\" ahler potential by adding the Dedekind function.

The scalar potential is
$$
V(T,\bar{T}) =
{{|j(T)|^{2}}\over{(T+\bar{T})^{3}|\eta(T)|^{12} }}
\left[{(T+\bar{T})^{2} \over 3}
 |{3 \over 2 \pi}\hat{G}_{2}(T,\bar{T})+{\partial_Tj(T) \over j(T)}|^{2}
 -3 \right].
\eqn\scalarpotential$$

The geodesic equation is  satisfied for
$T = e^{i\varphi(z)}$ ; \ie\ $T$ traverses the boundary of its
fundamental domain and it interpolates between the two isolated
supersymmetric minima, one at $T= \rho$
and another one at
$T=1$\refmark{\CFILQ}.
 (See figure 9 for the scalar potential
along the geodesic  $T=e^{i\varphi(z)}$.)

The proof that
the geodesic  corresponds
to $T=e^{i\varphi(z)}$  is straightforward.
First note $\partial_{T}j$ and $\hat G_2$ are both
modular forms of weight 2  while $j$ is the absolute
modular invariant  function. The results
$\partial_{T}j({1\over T})= -T^{2}\partial_{T}j(T)$,
$\hat G_2({1\over T})=
-T^2 \hat G_2(T)$, and
$j(e^{i\varphi}) =
\overline{ j(e^{i\varphi}) }$
imply
$ Im({D_TW\over W}\partial_{z}T)=0$
for $T=e^{i\varphi(z)}$ and thus satisfies the geodesic equation.

At the two  supersymmetric minima,
the superpotential takes values
$j(\rho) = 0$ and $j(1) = 1728$.
This in turn implies that the supersymmetric minima of the
potential  are non-degenerate.
At $T=1$ one has an anti-deSitter space with cosmological
constant $-3|W(T=1)|^2e^{K(T=1)}$  and at $T=e^{i\pi/6}$ the
cosmological constant is zero.
Even though the two supersymmetric
minima of the matter potential
are not degenerate (see figure 9 for the scalar potential along the
geodesic),
there does exist a stable  domain wall
solution interpolating between them.
Since $ie^{i\theta}={W\over |W|}=1$, $\zeta = -1$.
The coupled self-dual equations for $T=e^{i\varphi}$ and $A$ are
$$
\eqalign{
\partial_{z}\varphi(z) &=
\sqrt{A}K^{T \bar{T}}|W|e^{K\over 2}
{ \overline{ D_{T}W } \over iT\overline{W} }  \cr
\partial_{z}A(z) &= -2A^{3\over 2}|W|e^{K\over 2}
.}
\eqn\localselfdual$$
The results of numerical
integration of \localselfdual\ are shown in figures
10 and 11, respectively.
Since the potential does not have reflection
symmetry, we
 see that the metric
and the moduli fields are indeed not reflection symmetric.
\foot{As in the global example with the modular potential,
we found it necessary to scale the superpotential in order
to obtain these numerical results.  However, since the metric
function $A(z)$ is quite small, this function acts as a
scaling on the right hand side of \localselfdual\ .  Thus,
only a nominal value of $\Omega \approx 10^{-1}$ was
used in scaling the superpotential.}

Comparing this local example with the corresponding
global supersymmetric modular invariant theory, both cases
are similar; e.g.,
the two  isolated supersymmetric minima
are at $T=\rho$ and $T=1$ and the geodesic is the same
in both cases. However, a significant difference is that
in the local case the minima are not degenerate; \ie\ at $T=\rho$
the cosmological constant is zero, while at $T=1$ the
cosmological constant is negative.

\section{Comparisons between Local and Global Domain Walls}
The above two examples are representative  of a situation
where the study of the global supersymmetric domain wall
is readily generalizable to the local supersymmetric theory.
One may
be tempted to conclude that all the supersymmetric domain walls in
the global supersymmetric theory automatically remain as supersymmetric
domain wall solutions even after gravity is turned on.
However, this is not always the case. As we observed earlier
a criteria for the existence
of a domain wall is quite different in the two cases.
Here we comment on two typical examples where
 one obtains a supersymmetric
domain wall solution
 in the global case, whereas
there is {\it no}
analogous supersymmetric domain wall solution
  in the local case.

First consider another modular invariant superpotential:
$$
W  = j(T) (j(T) - 1728)
\eqn\newsuperpotential
$$
There are three isolated
global supersymmetric minima at
$T=1, \rho$ and $\partial W/\partial j = 2j(T)-1728 =0$.
Therefore, we expect two  domain walls
interpolating between each of the
two adjacent vacua.
In the supergravity case we find the minima
$T=1$ and $T=\rho$ remain supersymmetric minima. They both
have zero cosmological
constant since the
superpotential \newsuperpotential
 vanishes at these two points.\foot{ Note, again, that
$j(\rho)=0$ and $j(1)=1728$.}
Additionally, there
 is a local minimum with
positive cosmological constant at $T_{3}$ which is
in the neighborhood of the point
$j^{-1}(864) \in \cal D$.  However,
this point
is not supersymmetric since
$D_T W =[\partial_{T}W+{3\over{2\pi}}\hat G_2 W]|_{T=T_3}\ne 0$.
Thus, the domain wall  interpolating between $T=1$ and  $T_3$
(or between $T_3$ and $T= e^{i\pi/6}$) is not stable
since the minimum at $T_3$
is a non-supersymmetric \sl de-Sitter \rm minimuum.
Also, the
wall interpolating directly
 between the supersymmetric vacua at $T=1$ and $T=e^{i\pi/6}$
does not exist either as
the superpotential vanishes at these vacua and thus there
is no energy associated with such a wall.

Another example is a field theoretic case of a
$Z_3$ domain wall
 associated with the superpotential
$$W= {1\over 4} T^{4}-bT , \,\, b>0 \eqn\toysuperz$$
and a minimal K\"ahler potential $K=T\overline{T}$.
In the global case ~\refmark{\TOWN}
one has three isolated  minima at
$T=b^{1\over 3}e^{ 2\pi n i
\over 3}, \,\, n=0,1,2$. The geodesic lines in the $e^{K/2}W$-plane
(as well as in the $T$-plane) are
straight lines which are the sides of an equilateral triangle
interpolating among the three minima.
Note, however, that {\it none }  of the geodesic lines
lie along a   straight line that goes through the origin. On the other
hand, the global limit ($G_N\to 0$)   of the supergravity case
demands that the geodesic  should lie on a straight line that goes
through the origin in the $e^{K/2} W$-plane. This constraint cannot be
satisfied
    since
the minima are located in the $e^{K/2}W$
plane (as well as in the $T$ plane)
at the corners  of an equilateral triangle.
Thus, there is no domain wall solution when gravity is considered.

The above two
 examples clearly
show  that the criteria for the existence of a supersymmetric domain
wall in supergravity theories are quite constrained. These
constraints  are:
(1) The isolated vacua have to be
 supersymmetric,
 (2) The value of $W$ has to be non-zero
  at least at one of the two
 isolated supersymmetric vacua, (3) The corresponding
 global
 supersymmetric theory has to have  isolated minima lying  in the $e^{K/2} W$
plane along a straight line that extends through the origin,
(4) For the non-singular metric solution
 interpolating between the two
supersymmetric anti-deSitter vacua,
$W$ has to pass through zero.

\chap{Discussion}

We now provide some comments.
First, note that we obtained static
asymmetric domain walls.
These walls are not of the kind
studied in Ref.~\refmark{\IPSK}, where it was shown for
\sl infinitely thin,
reflection symmetric domain walls with asymptotic
Minkowski space-times on the both sides of the wall,
 the metric of
domain walls are generically time-dependent and not planar. \rm
Such assumptions and conclusions cover a general
class of matter sources such as a real scalar field theory with
double-well potential.
Here, in the study of domain walls in the supergravity theory,
we have observed that isolated
supersymmetric vacua, which need not be reflection symmetric
and at least one of them is anti-deSitter,
allow for static stable
domain wall solutions.
This is a class of domain wall solutions beyond those classified in
Ref.~\refmark{\IPSK}.

The constraints for the existence of supersymmetric domain walls
 in the supergravity theory were specified in the
previous chapter. Here we would like to highlight
the two most surprising results:
(1) There  are stable static domain wall solutions that
interpolate  between two isolated
but {\it non-degenerate} supersymmetric vacua.
This  leads us to define the notion of degeneracy of
vacua in supergravity theory  as those vacua that are
supersymmetric. Indeed, in supergravity, the total energy is a sum
of the matter and the gravitational energies. Therefore,
a proper notion
of the vacuum degeneracy should refer to the total energy,
not the matter energy alone.
(2) There  are examples of global supersymmetric
domain walls that do not have an analog when the
gravity is turned on; either the supersymmetry
is not preserved  at the minima anymore,
 or more interestingly, the global supersymmetric
theory has degenerate isolated minima which in the $e^{K/2}W$
plane do not lie on a straight line that extends through the
origin. This observation gives a clear message that
in the case of extended
topological defects, like domain walls,
 gravity plays a non-trivial and crucial role.

The domain walls considered   here
were those interpolating between
two non-degenerate
vacua of the supergravity matter potential, \eg\
one with zero and another with negative
cosmological constant.
The existence of such
static
domain walls is intimately related to the $O(4)$ symmetric
bubbles of the false vacuum decay.\Ref\COLEDEL{
S. Coleman
and F. DeLuccia, Phys. Rev. \bf D21 \rm (1980) 3305.}
In Ref.\refmark{\COLEDEL}
Coleman and DeLuccia found
that a false vacuum decay from the Minkowski space-time
to anti-deSitter space-time  cannot take place
\sl unless \rm the matter vacuum energy difference
$\epsilon \equiv V(false) - V(true)$ meets an inequality
$$
\epsilon \ge {3 \over 4 } \sigma^2
\eqn\coleman
$$
in which $\sigma$ denotes the energy density stored in the bubble wall.
The residual energy after materializing the bubble wall goes to
accelerate the wall asymptotically to the speed of light.
Also as the energy difference $\epsilon$ approaches the minimum of
the Coleman-DeLuccia
bound \coleman\ , $\ $ the radius of $O(4)$ invariant bubble wall becomes
indefinitely large. Precisely at the saturation limit,
$$\sigma = \sigma_c \equiv 2 \sqrt {\epsilon \over 3}. \eqn\crit $$
No kinetic energy
is available for the wall to accelerate to the speed of light, and the
wall radius becomes infinite, \ie\ becomes planar.
The resulting configuration of the $O(4)$ bubble is a
time-independent and infinite planar domain wall dividing
the Minkowski space-time from the anti-deSitter space-time.
In the
supergravity theory, however, $\sigma_c = 2 \sqrt {\epsilon \over 3}
= 2 e^{K/2} |W(false)|$ which coincides with the topological kink number
$|C|
= 2 |\Delta (\zeta e^{K/2} |W|)|$! Thus, the critical Coleman-DeLuccia
bubble
wall in supergravity theory is seen to saturate the Bogomolnyi bound
as well, hence,
this is a special class of the supersymmetric domain walls
we found in this paper.

The above
explanation also follows directly from the saturation of the positive
energy theorem\REF\WEIN{S. Weinberg, Phys. Rev. Lett. \bf 48 \rm
(1982) 1776.}\REF\OURS{M. Cveti\v c, S. Griffies and S.-J. Rey, \sl
Nonperturbative
Stability of Supersymmetric Vacua in $N=1$ Supergravity Theory \rm,
UPR-494-T, YCTP-P44-91
preprint
(December, 1991).} \refmark{\WEIN\ , \OURS}
for the false vacuum decay: the $O(4)$ bubble wall
energy density is bounded by the matter potential energy difference
and the minimum is saturated for a bubble tunneling from a supersymmetric
false vacua to another supersymmetric true vacua.
Details
of this proof and implications to superstring compactifications
are discussed elsewhere~\refmark{\OURS}.

We are grateful to G. Lopes-Cardoso, L.M. Krauss, V. Moncrief, D. Muraki,
P. Sikivie, M. Soldate and E. Weinberg for useful discussions.
M. C. and S. G.  are  supported in part by the
U.S. DOE Grant DE5-22418-281,
by the Grant from the Univ. of Pennsylvania
Research Foundation, by the NATO Research Grant
No. 900-700 and by the SSC fellowship.
S.-J. R. is supported in part by
the U.S. Department of Energy and
the Texas National Laboratory Research Commission.

\Appendix{A}

We present details necessary for evaluating the surface
and volume integrals leading to the inequality
in  the local theory
$$
\sigma \ge |C|.
\eqn\localbound$$

We start from eq.\localchargevariation\
$$
\delta_{\epsilon} Q[\epsilon']
= \int_{\partial \Sigma}N^{\mu \nu} d\Sigma_{\mu \nu}
= 2\int_{\Sigma}\nabla_{\nu}N^{\mu \nu} d\Sigma_{\mu}.
\eqn\localchargevariationappendix$$

This equation makes essential use of Stoke's Law and the
covariance of the generalized Nester's Form.  The strategy
is to extract information about both the ADM mass density of the
wall in its rest frame and the topological charge
from the surface integral.  The volume integral will
be shown positive semidefinite
by making certain assumptions about the Majorana
spinors $\epsilon, \epsilon'$.  Saturation of the inequality will
occur for supersymmetric bosonic backgrounds which then lead
to first order differential equations (the self-dual equations)
for the matter field $T$ and space-time metric.
Solutions of these first order equations can easily be shown to
satisfy the usual second order equations obtained from the
Lagrangian \localL\ .  Solving these equations will yield
(1) an explicit expression for the ADM mass density of the
supersymmetric domain wall, (2) the space-time geometry
and (3) the matter field configuration.
We note that the spinors $\epsilon,\epsilon'$ are introduced
to obtain information about the ADM mass density and
topological charge density of the system.
A priori they are constrained only to satisfy
$\hat{\nabla}_{\mu}\epsilon \to O(z^{-1})$ as $|z| \to \infty$;
\ie\ they are asymptotic Killing spinors of the
bosonic background~\refmark{\GHW} .
For supersymmetric backgrounds,
$\hat{\nabla}_{\mu}\epsilon \equiv 0$, which implies they must
satisfy a particluar first order differential equation and thus
loose their arbitrariness; they become Killing spinors.
As in the global case, only two Killing spinors exist
for the supersymmetric domain wall background.

We are concerned with supercharge \sl density \rm and thus
insist upon only $SO(1,1)$ covariance, where the spatial direction
is transverse to the wall ($\hat{z}$).  We use the space-time
metric
$g_{\mu \nu} = diag(A,-B,-B,-A)$
and choose the veirbein
$e_{\mu}^{a} = diag(A^{1\over2},B^{1\over2},B^{1\over2},A^{1\over2})$.
$A$ and $B$ are functions of $z,t$.
This prescription implies
the spacelike hypersurface
$\Sigma$ is just the $z-$axis
with measure
$d\Sigma_{\mu} = (d\Sigma_{t} =
|g_{tt}g_{zz}|^{1\over 2}dz, 0,0,0)$.  The boundary
$\partial \Sigma$ is the
two points $z\rightarrow \pm\infty$.

Making use of the result valid for an
antisymmetric tensor
(remember the $SO(1,1)$ covariance)
$$
\nabla_{\nu}N^{\mu \nu} =
|g_{tt}g_{zz}|^{-1/2}\partial_{\nu}[|g_{tt}g_{zz}|^{1/2}N^{\mu \nu}]
\eqn\genreltrick$$
and assuming only $z$-dependence
allows us to write \localchargevariationappendix\
as
$$
\delta_{\epsilon} Q[\epsilon'] =  2\int_{-\infty}^{\infty}
|g_{tt}g_{zz}|^{-1/2}\partial_{\nu}(|g_{tt}g_{zz}|^{1/2}N^{\mu \nu})
|g_{tt}g_{zz}|^{1/2}dz =
2\Delta(|g_{tt}g_{zz}|^{1\over 2}N^{tz}).
\eqn\surfaceterm$$

Nester's form involves three terms:
$$
N^{tz} =
\overline{\epsilon'}\gamma^{tz\rho}
(2\nabla_{\rho})\epsilon
+ \overline{\epsilon'}\gamma^{tz\rho}
ie^{K \over 2}(WP_{R} + \bar{W}P_{L})\gamma_{\rho}\epsilon
- \overline{\epsilon'}\gamma^{5}Im(K_{T}\partial_{\rho}T)\epsilon.
\eqn\nerstertz$$
The first term
involves the gravitational covariant derivative $\nabla_{\rho}$
and yields the
ADM mass density of the configuration~\refmark{\WITTEN,\GHW} which
we denote as $\sigma$.

As $T=T(z)$ and
$\gamma^{\mu \nu \rho} = \gamma^{[\mu}\gamma^{\nu}\gamma^{\rho]}$,
the $Im(K_{T}\partial_{\rho}T)$ term drops out.
The second piece yields the
topological term
of Nester's form.  We use the identities
$\gamma^{tz\rho}\gamma_{\rho} =
4\sigma^{tz} = 2\gamma^{t}\gamma^{z} = 2A^{-1}\gamma^{0}\gamma^{3}$,
$\overline{\epsilon} = \epsilon^{\dagger}\gamma^{t} =
\epsilon^{\dagger}\gamma^{0}A^{-1/2}$ and
$|g_{tt}g_{zz}|^{1/2} =A$  to find
$$
2N^{tz}_{top}|g_{tt}g_{zz}|^{1/2} =
4iA^{-1/2}
\epsilon'^{\dagger}\gamma^{3}e^{K\over2}(ReW + \gamma^{5}ImW)\epsilon
\equiv C(z).\eqn\topological$$
The factor of $A^{-1/2}$ will drop out for the
supersymmetric background case
since the spinors
will have an $A^{1/4}$ dependence (see Appendix B).

In summary, we have for the surface integral  of Nester's form
$$
2\Delta[|g_{tt}g_{zz}|^{1\over 2}N^{tz}] =
2(\sigma + C)
\eqn\summarynester$$
where $\Delta$ denotes the value of its argument evaluated
at $z=+\infty$ minus that at $z=-\infty$.

Recall the
only constraint we impose on the
spinors is that they satisfy $\hat{\nabla}_{\mu}\epsilon = O(z^{-1})$
as $|z| \to \infty$.  For static supersymmetric bosonic backgrounds
these spinors will
satisfy $\hat{\nabla}_{\mu}\epsilon = 0$ and an explicit
expression for $\sigma = |C|$ will be found (see Appendix B).

We now proceed to show that the volume integral is positive
definite. For that purpose
the gravitational covariant derivative of Nester's form can be written
as:
$$
\eqalign{
2\nabla_{\nu}N^{\mu \nu}
&= 2\overline{ \nabla_{\nu}\epsilon'}
\gamma^{\mu \nu \sigma} \hat{\nabla}_{\sigma}\epsilon
+
2\overline{ \epsilon' }\gamma^{\mu \nu \sigma}
\nabla_{\nu}\hat{\nabla}_{\sigma}\epsilon  \cr
&=
\overline{ \hat{\nabla}_{\nu}\epsilon' }
\gamma^{\mu \nu \sigma} \hat{\nabla}_{\sigma}\epsilon
+
\overline{ \epsilon' }\gamma^{\mu \nu \sigma}
\hat{\nabla_{\nu} }\hat{\nabla}_{\sigma}\epsilon  \cr
&+
\overline{ (-ie^{K\over2}(WP_{R}+\overline{W}P_{L})\gamma_{\nu}
+ Im(K_{T}\partial_{\nu}T)
\gamma^{5})\epsilon' } \gamma^{\mu \nu \sigma}
\hat{\nabla}_{\sigma}\epsilon  \cr
&+
\overline{ \epsilon' } \gamma^{\mu \nu \sigma}
(-ie^{K\over2}(WP_{R}+\overline{W}P_{L})\gamma_{\nu}
+ Im(K_{T}\partial_{\nu}T)\gamma^{5})
\hat{\nabla}_{\sigma}\epsilon.
}
\eqn\nesterdiv$$
The last two terms in
\nesterdiv\ cancel identically.
Using Dirac algebra and standard properties of
gravitational derivatives yields:
$$
\eqalign{
\overline{ \epsilon' }\gamma^{\mu \nu \sigma}
\hat{\nabla_{\nu}}\hat{\nabla}_{\sigma}\epsilon
&=
{1\over2} \overline{ \epsilon' }\gamma^{\mu \nu \sigma}
[\hat{\nabla}_{\nu},\hat{\nabla}_{\sigma}]\epsilon \cr
&=
2ie^{K\over2}(\overline{\epsilon'}
\gamma^{\mu \nu \sigma}\gamma^{5}\epsilon)
K_{T\overline{T}}\partial_{\nu}\bar{T}\partial_{\sigma}T
+
6|We^{K\over 2}|^{2}
(\overline{ \epsilon' }\gamma^{\mu}\epsilon) \cr
+
2(\overline{\epsilon'}\gamma^{\nu}\epsilon)G^{\mu}_{\nu}
&+
8ie^{K \over2}
\overline{\epsilon'}\sigma^{\mu \nu}
[Re(D_{T}W\partial_{\nu}T)+\gamma^{5}Im(D_{T}W\partial_{\nu}T)]
\epsilon
}
\eqn\nestermore$$
where $G^{\mu \nu} = R^{\mu \nu} - {1 \over2}Rg^{\mu \nu}$,
$\sigma^{\mu \nu} = {1 \over4}[\gamma^{\mu},
\gamma^{\nu}]$, and $[\nabla_{\nu},\nabla_{\mu}]\epsilon = {1\over2}
R^{ab}_{\mu \nu}\sigma_{ab}\epsilon$.

We can simplify this expression  by introducing the bilinear
$\Gamma^{\mu} = K_{ T \bar{T} }\overline{\delta_{\epsilon'}\chi}
\gamma^{\mu}\delta_{\epsilon}\chi$ as in the global case.\foot{See
eq. \formin\
and
the following footnote regarding commuting vs. anti-commuting spinors.}
The variation of the spin $1/2$ field $\chi$ in $N=1$ supergravity is
$$
\delta_{\epsilon}\chi = -\sqrt2 e^{K\over2}K^{T \bar T}
(D_{\overline{T}}\overline{W}P_{R} +
D_{T}WP_{L})\epsilon
- i\sqrt2 (\partial_{\nu}T P_{R} + \partial_{\nu}\bar T P_{L})
 \gamma^{\nu} \epsilon,
\eqn\modulinolocal$$
and the matter energy-momentum tensor is
$$
T^{\mu \nu} = -g^{\mu \nu}[K_{T \bar{T}} \partial_{\sigma}T
\partial^{\sigma}\bar{T} - V(T,\bar{T})]
+ K_{T \bar{T} }(\partial^{\nu}\bar T\partial^{\mu}T + \partial^{\nu}T
\partial^{\mu}\bar{T}),
\eqn\energytensor$$  where $V$ is the scalar potential(cf. eq.
\localL\ ).
Using eqs.\energytensor\ in $\Gamma^{\mu}$ and
$\Gamma^{\mu}$ in \nestermore\ yields
$$
2\nabla_{\nu}N^{\mu \nu}
=
\overline{ \hat{\nabla}_{\nu}\epsilon' }
\gamma^{\mu \nu \sigma} \hat{\nabla}_{\sigma}\epsilon
+
\Gamma^{\mu}
+
2\overline{\epsilon'}\gamma_{\nu}\epsilon
(G^{\mu \nu} - T^{\mu \nu }).
\eqn\nesterdiverge$$
The last term vanishes upon imposing Einstein's equation.

To integrate on a space-like hypersurface, we need
$$
\nabla_{\nu}N^{t\nu} =
A^{-1}(\gamma^{i}\hat{\nabla_{i}}\epsilon)^{\dagger}
(\gamma^{j}\hat{\nabla_{j}}\epsilon)
- A^{-1}g^{ij}(\hat{\nabla_{i}}\epsilon)^{\dagger}
(\hat{\nabla_{j}}\epsilon)
+ \Gamma^{t}
\eqn\nesterdivergeagain$$  where the vierbein factor
follows from $\gamma^{t} = A^{-1/2}\gamma^{0}$.
In order to obtain an inequality,
we set $\epsilon = \epsilon^{\prime}$.  Also note $i,j = x,y,z$.
For commuting Majorana
spinors, $(+,-,-,-)$ signature, and $A>0$,  we know $\Gamma^{t}
- A^{-1}g^{ij}(\hat{\nabla_{i}}\epsilon)^{\dagger}
  (\hat{\nabla_{j}}\epsilon) \ge 0$.  Imposing the
generalized Witten condition
$\gamma^{i}\hat{\nabla_{i}}\epsilon = 0$ ensures the bound
$\nabla_{\nu}N^{t\nu} \ge 0$.

In terms of the supersymmetry variations of the fermion
fields, we have
$$
2A\nabla_{\nu}N^{t \nu}
=
 -\delta_{\epsilon}\psi_{i}^{\dagger}g^{ij}
\delta_{\epsilon}\psi_{j}
+
K_{ T \bar{T} }\delta_{\epsilon}\chi^{\dagger}
\delta_{\epsilon}\chi  \ge 0.
\eqn\nesterdivergeagain$$

Combining the results for the surface and volume integral
yields the inequality
$$
\delta_{\epsilon} Q[\epsilon]
\equiv \{Q[\epsilon],
\bar{Q}[\epsilon]\}  =
2(\sigma + C) \ge 0.
\eqn\localbound$$
For supersymmetric configurations
($\delta_{\epsilon} Q[\epsilon] \equiv 0$)
we solve the self-dual equations (see Appendix B) and find
$$
\sigma
= |C| =
2|\Delta(\zeta|We^{K\over2}|)|. \eqn\admmass$$
Note the topological charge $C$ is real. This
is because gravity requires $\epsilon$ to satisfy a particular
differential equation (see Appendix B).
Therefore, the object $C(z)$ must be evaluated at each
infinity, including the multiplication of the spinor
$\epsilon$.
Using results of Appendix B, we can understand these statements
clearly.  The topological term is
$$
C(z) \equiv AN^{tz}(z)_{topological} =
2iA^{-1/2}e^{K \over 2}\epsilon^{\dagger}
(\gamma^{3}ReW + \gamma^{3}\gamma^{5}ImW)\epsilon.
\eqn\topologicalcharge$$
The topological charge is the difference of $C(z)$ at the
two infinities.  Multiplying through by the known $\epsilon$
yields
$C =  2\Delta(\zeta|We^{K \over 2}|)$.

The supercharge density algebra \localbound\ is an $N=2$
supersymmetry algebra.  Recall we are working in the wall's rest
frame.  Therefore, the $N=2$ algebra is the same for
asymptotic
Minkowski or anti-deSitter space-times~\refmark{\GHW}.

\Appendix{B}

In this Appendix we derive the self-dual equations for the
space-time metric components $A,B$ from imposing
$\hat{\nabla}_{\mu}\epsilon = 0$.
In addition, we derive and solve the differential equation
for the spinor $\epsilon$ introduced in Nester's form.

First we exhibit the supercovariant derivative
$\hat{\nabla}$ appearing in the supersymmetry variation
of the gravitino: $\delta_{\epsilon}\psi_{\mu} =
\hat{\nabla}_{\mu}\epsilon$.
The space-time metric is $g_{\mu \nu} = diag(A,-B,-B,-A)$
and we choose the veirbein
$e_{\mu}^{a} = diag(A^{1\over2},B^{1\over2},B^{1\over2},A^{1\over2})$.
$A$ and $B$ are functions of $z,t$.
Construction of the spin connection $\omega^{ab}_{\mu}$
is a straightforward relativity
exercise~\refmark{\RELA}.
We find
$$
\eqalign{
\hat{\nabla}_{t}\epsilon &= \left[ 2\partial_{t}
+ {1\over 2}\partial_{z}lnA\gamma^{0}\gamma^{3}
+ i\sqrt{A}\gamma^{0}(WP_{L}+\overline{W}P_{R})e^{K\over 2}
\right]\epsilon \cr
\hat{\nabla}_{x}\epsilon &= \left[ 2\partial_{x}
+ {1\over 2}{\left({B\over A}\right)}^{1/2}
\partial_{t}lnB\gamma^{0}\gamma^{1}
- {1\over 2}{\left({B\over A}\right)}^{1/2}\partial_{z}lnB
\gamma^{1}\gamma^{3}
- i\sqrt{B}\gamma^{1}(WP_{L}+\overline{W}P_{R})e^{K\over 2}
\right]\epsilon \cr
\hat{\nabla}_{y}\epsilon &= \left[ 2\partial_{y}
+ {1\over 2}{\left({B\over A}\right)}^{1/2}\partial_{t}lnB\gamma^{0}
\gamma^{2}
- {1\over 2}{\left({B\over A}\right)}^{1/2}
\partial_{z}lnB\gamma^{2}\gamma^{3}
- i\sqrt{B}\gamma^{2}(WP_{L}+\overline{W}P_{R})e^{K\over 2}
\right]\epsilon \cr
\hat{\nabla}_{z}\epsilon &= \left[ 2\partial_{z}
+ {1\over 2}\partial_{t}lnA\gamma^{0}\gamma^{3}
- i\sqrt{A}\gamma^{3}(WP_{L}+\overline{W}P_{R})e^{K\over 2}
- \gamma^{5}Im(K_{T}\partial_{z}T)\right]\epsilon
}
\eqn\supercov$$
We take the Majorana spinor $\epsilon$ in the form
$\epsilon=(\epsilon_1,\epsilon_2,\epsilon_2^*,-\epsilon_1^*)$
which is consistent with our using a Weyl basis for the
flat Dirac matrices.  Recall
from setting $\delta_{\epsilon}\chi = 0$ that
$\epsilon_1=e^{i\theta}\epsilon_2^*$.

Taking the Ansatz of time independent metric
components $A,B$
yields
$$
\partial_{z}lnA =
2(ie^{-i\theta})\sqrt{A}e^{K\over 2}W
\eqn\Aequ$$
from
$\hat{\nabla}_{t}\epsilon = 0$,
$$
\partial_{z}lnB =
2(ie^{-i\theta})\sqrt{A}e^{K\over 2}W
\eqn\Bequ$$ from
$\hat{\nabla}_{x}\epsilon = 0$ \&
$\hat{\nabla}_{y}\epsilon = 0$, and
$$
 \eqalign{
[2\partial_{z} + iIm(K_{T}\partial_{z}T)](\epsilon^{*}_{2}e^{i\theta})
 &=
{1\over 2}(\partial_{z}lnA)\epsilon^{*}_{2}e^{i\theta} \cr
[2\partial_{z} + iIm(K_{T}\partial_{z}T)]\epsilon_{2}
 &=
{1\over 2}(\partial_{z}lnA)\epsilon_{2}
 }
\eqn\epsilonequ$$ from
$\hat{\nabla}_{z}\epsilon = 0$.
Eq. \epsilonequ\ implies the phase $\theta$ must satisfy
$$\partial_{z}\theta = -Im(K_{T}\partial_{z}T).
\eqn\thetaequ$$
We can solve for
$\epsilon_{2}$ from \epsilonequ\ to find
$$
\epsilon_{2}(z) = {1\over2}A^{1\over 4}(z)
e^{-i/2 \int^{z}_{c} \!\!dz'\,\,Im(K_{T}\partial_{z}T) }
=
{1\over2}A^{1\over 4}(z)e^{{1 \over 2} i\theta(z)}
\eqn\epsilonsolved$$
where we normalized $\epsilon(z)$ such that
$\overline{\epsilon}\gamma^{0}\epsilon \equiv
(\epsilon^{\dagger}A^{-1/2}\gamma^{0})\gamma^{0}\epsilon = 1$.
Notice the important $A^{1\over4}$ behavior( cf. eq. \topological\ )
as well as the
$z-$dependent phase.
\endpage
\refout
\endpage

{\bf Figure Captions}

\
\

\noindent{
Figure 1: Global double-well potential with
$a^{2}=0.5$
along the geodesic $T(z) \in \bf{R}$.
}

\
\

\noindent{
Figure 2: Global domain wall $T(z) \in \bf{R}$
for double-well potential with $a^{2}=0.5$.
}

\
\

\noindent{
Figure 3: Fundamental domain for $PSL(2,{\bf Z})$.
}

\
\

\noindent{
Figure 4: Global modular invariant potential along the geodesic
$T=e^{i\varphi(z)}$.
}
\break
\noindent{
Superpotential scaled by $\Omega = 2 \times 10^{-5}$.
}

\
\

\noindent{
Figure 5: Global domain wall $T(z)=e^{i\varphi(z)}$ for modular
covariant potential.
}
\break
\noindent{
Superpotential scaled by $\Omega = 2 \times 10^{-5}$.
}

\
\

\noindent{
Figure 6: Local double-well potential for $a^{2}=.5$.
}

\
\

\noindent{
Figure 7: Local domain wall
$T(z) \in \bf{R}$ for double-well potential.
}

\
\

\noindent{
Figure 8: Metric function $A(z)$ for double-well potential.
}

\
\

\noindent{
Figure 9: Local modular invariant potential along the geodesic
$T(z)=e^{i\varphi(z)}$.
}
\break
\noindent{
Superpotential scaled by $\Omega = 3 \times 10^{-1}$.
}

\
\

\noindent{
Figure 10: Local domain wall $T(z)=e^{i\varphi(z)}$ for modular
invariant potential.
}
\break
\noindent{
Superpotential scaled by $\Omega = 3 \times 10^{-1}$.
}

\
\

\noindent{
Figure 11: Metric function $A(z)$ for modular invariant potential.
}
\break
\noindent{
Superpotential scaled by $\Omega = 3 \times 10^{-1}$.
}

\end